\documentclass[%
11pt,
nofootinbib,
notitlepage,
amsfonts,
amsmath,
amssymb,
twocolumn,
pra,
superscriptadress,
longbibliography
]{revtex4-1}

\usepackage{mathptmx}
\usepackage{graphicx}
\usepackage{bm}
\usepackage{braket}
\usepackage{xcolor}
\usepackage{mdwlist}

\usepackage[top=22truemm,bottom=22truemm,left=23truemm,right=23truemm]{geometry}

\begin{document}


\title{{High-threshold fault-tolerant quantum computation with analog quantum error correction}}

\author{Kosuke Fukui}
\author{Akihisa Tomita}%
\author{Atsushi Okamoto}%
\affiliation{%
Graduate School of Information Science and Technology, \\Hokkaido University, Kita14-Nishi9, Kita-ku, Sapporo 060-0814, Japan}
\author{Keisuke Fujii}
\affiliation{%
Department of Physics, Graduate School of Science, Kyoto University, \\Kitashirakawa-Oiwakecho, Sakyo-ku, Kyoto 606-8502, Japan}

\begin{abstract}
To implement fault-tolerant quantum computation with continuous variables, the Gottesman--Kitaev--Preskill (GKP) qubit has been recognized as an important technological element. However, it is still challenging to experimentally generate the GKP qubit with the required squeezing level, 14.8 dB, of the existing fault-tolerant quantum computation. To reduce this requirement, we propose a high-threshold fault-tolerant quantum computation with GKP qubits using topologically protected measurement-based quantum computation with the surface code. By harnessing analog information contained in the GKP qubits, we apply analog quantum error correction to the surface code. Furthermore, we develop a method to prevent the squeezing level from decreasing during the construction of the large scale cluster states for the topologically protected measurement based quantum computation. We numerically show that the required squeezing level can be relaxed to less than 10 dB, which is within the reach of the current experimental technology. Hence, this work can considerably alleviate this experimental requirement and take a step closer to the realization of large scale quantum computation.
\end{abstract}
\maketitle
\section{Introduction}\label{sec1}
Quantum computation has a great deal of potential to efficiently solve some hard problems for conventional computers~\cite{Shor,Grov}. Although small-scale quantum computation with various phycsical systems has been demonstrated, large-scale quantum computation is still a significant experimental challenge for most candidates of physical systems. Among the candidates, squeezed vacuum states in an optical system have shown great potential for large scale continuous variable (CV) quantum computation; in fact, more than one million-mode CV cluster state has been achieved already in an experiment~\cite{Yoshi}. This ability of entanglement generation comes from the fact that squeezed vacuum states can be entangled  by using the time-domain multiplexing approach by only beam splitter coupling to miniaturize optical circuits~\cite{Meni4,Meni5}. 

Since CV quantum computation itself has an analog nature, it is difficult to handle the accumulation of analog errors caused, for example, by photon loss during quantum computation~\cite{Ohl,Ohl2}. This can be circumvented by encoding digitized variables into CVs using an appropriate code, such as Gottesman--Kitaev--Preskill (GKP) code~\cite{GKP}, which are referred to as GKP qubits. By digitizing CVs, the standard quantum error correcting (QEC) code can be applied to implement fault-tolerant quantum computation (FTQC) with CVs. Moreover, GKP qubits inherit the advantage of squeezed vacuum states on optical implementation; they can be entangled by only beam splitter coupling. Furthermore, qubit-level Clifford gates on the GKP qubits in measurement-based quantum computation (MBQC) are implemented by Gaussian operation achieved simply by a homodyne measurement on CV cluster states~\cite{Meni3}. Menicucci showed that CV-FTQC is possible within the framework of MBQC using squeezed vacuum cluster states with GKP qubits~\cite{Meni}. A promising architecture for a scalable quantum circuit has been proposed recently~\cite{Tak,Ale}, where the GKP qubits are incorporated to perform FTQC. Hence, the GKP qubits will play an indispensable role in implementing CV-FTQC. 

Regarding the generation of the GKP qubit, a promising proposal~\cite{Ter} exists to prepare a good GKP qubit in circuit quantum electrodynamics with the squeezing level around 10 dB~\cite{Cas} within the reach of near-term experimental set-up. This implies that large scale quantum computation is possible, if the required squeezing level of the initial single qubit for FTQC is less than 10 dB. Yet, there is a large gap between the experimentally achievable squeezing level and theoretical requirement squeezing level. For example, the existing CV-FTQC requires the squeezing level of both squeezed vacuum state and GKP qubit 14.8--20.5 dB~\cite{Meni} to achieve the fault-tolerant threshold $2 \times 10^{-2}$~\cite{Knill,Fuji1,Fuji2} $- 10^{-6}$~\cite{Knill2,Pre,Kitaev}. Therefore, it is highly desirable to reduce the required squeezing level to around 10 dB to realize the large scale CV-FTQC.

In this work, we propose a high-threshold FTQC to alleviate the required squeezing level for FTQC by harnessing analog information contained in the GKP qubit. The analog information obtained by measuring CV states (including GKP qubits)  reflects the effect of noise as a deviation in the measurement outcome. Therefore, it contains beneficial information to improve the error tolerance. The proposed high-threshold FTQC consists of two parts. One is to apply analog QEC~\cite{KF1} to the surface code, which allows us to implement the high-threshold FTQC. The other is a construction of the cluster state for topologically protected MBQC \cite{Kitaev2,Rau,Rau2,Rau3} with a low error accumulation by using the postselected measurement with the help of analog information. In general, the accumulation of errors on a qubit, which causes degradation of the threshold, increases as the number of the entangling gate increases. In this work, we develop a novel method to avoid this accumulation of errors by using the proposed postselected measurement which harnesses analog information. Accordingly, the required squeezing level for topologically protected MBQC with the 3D cluster state constructed by our method can be reduced to 9.8 dB. By contrast, the required squeezing level in the existing CV-FTQC scheme~\cite{Meni} combined with the fault-tolerant scheme with the threshold $0.67\times 10^{-2}$~\cite{Rau3} is 16.0 dB~\cite{Note}. This improvement results from the reduction from 16.0 dB to 9.8 dB corresponds to the reduction of the error probability to misidentify the single GKP qubit in $q$ and $p$ quadrature from $ 2.7\times 10^{-15}$ to $7.4\times 10^{-5}$. By achieving the requirement of the squeezing level around 10 dB, we believe this work can considerably take a step closer to the realization of large-scale quantum computation with digitized CV states and will be indispensable to construct CV-FTQC.

The rest of the paper is organized as follows. In Sec. \ref{sec2} we briefly review the GKP qubit and the analog QEC, and apply the analog QEC to a surface code. In Sec. \ref{sec3} we propose the postselected measurement and present a high-threshold FTQC on the 3D cluster state constructed by using the postselected measurement. In Sec. \ref{sec4} the required squeezing level is calculated. We first calculate the unheralded error in the leading order for simplicity and then we simulate the analog QEC on the 3D cluster states constructed by using the fusion gate with the postselected measurement. Section \ref{sec5} is devoted to a discussion and conclusion.
\section{Analog quantum error correction}\label{sec2}
\subsection{The GKP qubit}
We review the GKP qubit and the error model considered in this work. Gottesman, Kitaev, and Preskill proposed a method to encode a qubit in an oscillator's $q$ (position) and $p$ (momentum) quadratures to correct errors caused by a small deviation in the $q$ and $p$ quadratures. This error correction of a small deviation can handle any error acting on the oscillator, which can be expanded as a superposition of displacements~\cite{GKP}. The basis of the GKP qubit is composed of a series of Gaussian peaks of width $\sigma$ and separation $\sqrt{\pi}$ embedded in a larger Gaussian envelope of width 1/$\sigma$. Although in the case of infinite squeezing ($\sigma \rightarrow 0$) the GKP qubit bases become orthogonal, in the case of finite squeezing, the approximate code states are not orthogonal. The approximate code states $\ket {\widetilde{0}}$ and $\ket {\widetilde{1}}$ is defined as  
\begin{eqnarray}
\ket {\widetilde{0}} \propto  \sum_{t=- \infty}^{\infty} \int \mathrm{e}^{-2\pi\sigma^2t^2}\mathrm{e}^{-(q-2t\sqrt{\pi})^2/(2\sigma^2)}\ket{q}  dq,    \\ 
\ket {\widetilde{1}} \propto  \sum_{t=- \infty}^{\infty} \int \mathrm{e}^{-\pi\sigma^2(2t+1)^2/2} \hspace{80pt} \nonumber \\ 
\mathrm{e}^{-(q-(2t+1)\sqrt{\pi})^2/(2\sigma^2)}\ket{q}  dq .     
\end{eqnarray}
In the case of finite squeezing there is a probability of misidentifying $\ket {\widetilde{0}}$ as $\ket {\widetilde{1}}$, and vice versa. Provided the magnitude of the true deviation is less than $\sqrt{\pi}/2$ from the peak value, the decision of the bit value from the measurement of the GKP qubit is correct. The probability $p_{\rm corr}$ to identify the correct bit value is the area of a normalized Gaussian of a variance ${{\sigma}}^2$ that lies between $-\sqrt{\pi}/2$ and $\sqrt{\pi}/2$~\cite{Meni}:
\begin{equation}
p_{\rm corr} = \int_{\frac{-\sqrt{\pi}}{2}}^{\frac{\sqrt{\pi}}{2}} dx \frac{1}{\sqrt{2\pi {\sigma} ^2}} {\rm exp}(-x^2/{2{\sigma} ^2}).
\end{equation}
In addition to the imperfection that originates from the finite squeezing of the initial states, we consider the Gaussian quantum channel~\cite{GKP,Harri}, which leads to a displacement in the quadrature during the quantum computation. The channel is described by superoperator $\zeta $ acting on density operator $\rho$ as follows: 
\begin{equation}
\rho \to \zeta (\rho) = \frac{1}{\pi{\xi }^2}\int d^2\alpha \mathrm{e}^{-{| \alpha |}^2/{{\xi}^2}}D( \alpha ) \rho D( \alpha ) ^{\dagger },
\end{equation}
where $D(\alpha)$ is a displacement operator in the phase space. The position $q$ and momentum $p$ are displaced independently as follows:
\begin{equation}
 q \to q + v , \  \ p \to p + u,
 \end{equation}
where $v$ and $u$ are real Gaussian random variables with mean zero and variance $\xi ^2$. The Gaussian quantum channel conserves the position of the Gaussian peaks in the probability density function on the measurement outcome of the GKP qubit, but increases the variance as follows:
\begin{equation}
\sigma^2  \to \sigma^2 \ + \xi^2,
\end{equation}
where the $\sigma^2$ is the variance before the Gaussian quantum channel. Therefore, in the next section, we evaluate the performance under a code capacity noise model, where the noise is parameterized by a single variance $\sigma^2$ that includes the squeezing level of the initial GKP qubit and the degradation via the Gaussian quantum channel.
\subsection{Analog quantum error correction}
\begin{figure}[t]
 \includegraphics[angle=270, width=1.0\columnwidth]{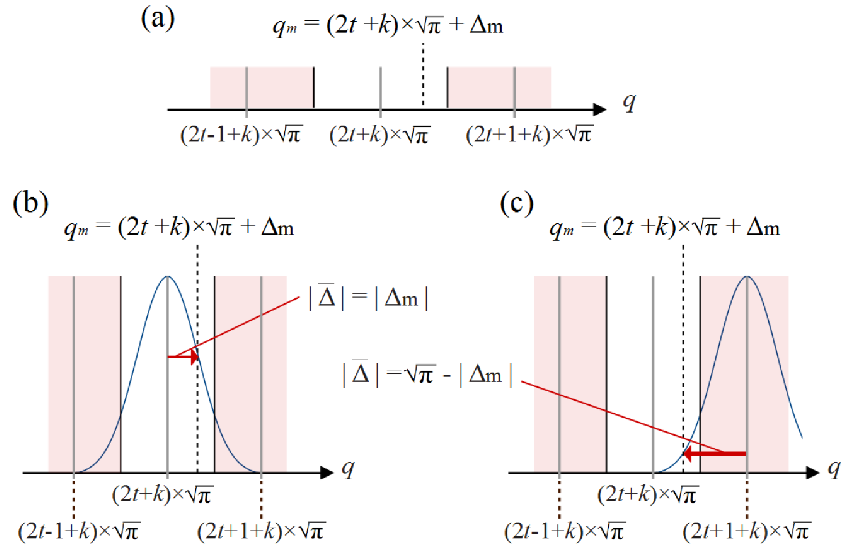}
     \setlength\abovecaptionskip{8pt}
 \caption{\label{Fig1}Introduction of a likelihood function. (a) Measurement outcome and deviation from the peak value in $q$ quadrature. The dotted line shows the measurement outcome $q_{\rm m}$ equal to $(2 t + k)\sqrt{\pi}+\Delta_{\rm m}$ $(t = 0, \pm 1, \pm 2,\cdots,\ k = 0, 1)$, where $k$ is defined as the bit value that minimizes the deviation $\Delta_{\rm m}$. The red areas indicate the area that yields code word $(k+1)$ mod 2, whereas the white area denotes the area that yields the codeword $k$. (b) and (c) Gaussian distribution functions as likelihood functions of the true deviation $\overline{\Delta}$ represented by the arrows. (b) refers to the case of the correct decision, where the amplitude of the true deviation is $|\overline{\Delta}| < \sqrt{\pi}/2$, whereas (c)  the case of the incorrect decision $ \sqrt{\pi}/2 < |\overline{\Delta}| < \sqrt{\pi}$.}
 \label{fig:1}
\end{figure}
Before describing our proposal using the surface code, we explain how the analog QEC works in general to harness analog information contained in the GKP qubit to improve the performance of the QEC (see also~\cite{KF1} for details). In the measurement of the GKP qubit for the computational basis, we make a decision on the bit value $k ( = 0, 1)$ from the measurement outcome of the GKP qubit in the $q$ quadrature $q_{\rm m}= q_{k}+ \Delta_{\rm m}$ to minimize the deviation $|\Delta_{\rm m}|$, where $q_{k}(k = 0,1)$ is defined as $(2 t + k)\sqrt{\pi} (t = 0, \pm 1, \pm 2,\cdots.)$ as shown in Fig. \ref{fig:1} (a). In the digital QEC~\cite{Pou,Goto}, we obtain the bit value from analog outcome and calculate the likelihood from only binary information, since we consider the GKP qubit as a qubit. The likelihood of the correct decision in Fig. \ref{fig:1} (b) using only bit value, and the noise level ${\sigma} ^2$ is calculated by
\begin{equation}
 p_{\rm corr} = \int_{\frac{-\sqrt{\pi}}{2}}^{\frac{\sqrt{\pi}}{2}} dx \frac{1}{\sqrt{2\pi {\sigma} ^2}} {\rm exp}(-x^2/{2{\sigma} ^2}). 
 \label{eq:7}
\end{equation}
The likelihood of the incorrect decision in Fig. \ref{fig:1} (c) is calculated by $1- p_{\rm corr}$. In the analog QEC, we employ Gaussian function, which the true deviation $|\overline{\Delta}|$ obeys, as likelihood function and calculate the likelihoods for decision of the bit value. The likelihood of the correct decision is calculated by
\begin{equation}
 f(\overline{\Delta}) = f(\Delta_{\rm m}) = \frac{1}{\sqrt{2\pi\sigma^{2}}} \mathrm{e}^{-\overline{\Delta}^{2}/(2\sigma^{2})}. 
  \label{eq:8}
\end{equation}
The likelihood of the incorrect decision is calculated by 
\begin{equation}
f(\overline{\Delta}) =f(\sqrt{\pi}-|\Delta_{\rm m}|).
 \label{eq:9}
\end{equation}
Strictly speaking, the likelihood function should be the periodic function including the sum of the Gaussian functions, considering the superposition of the Gaussian states. In this paper, the likelihood function is approximated by simple Gaussian functions given by Eqs. (\ref{eq:8}) and (\ref{eq:9}), since the tail of the Gaussian function next to the measurement outcome is small enough to ignore~\cite{Note2}. In the QEC, we can reduce the decision error on the entire code word by considering the likelihood of the joint event of multiple qubits and choosing the most likely candidate. The analog QEC under the code capacity model can improve the QEC performance with a single block code without the concatenation such as the three-qubit flip code~\cite{KF1}. In the previous proposal based on the digital QEC~\cite{Pou,Goto} can improve the QEC performance with only the concatenated code under the code capacity model, since the improvement results from the message-passing algorithm. Under the several noise models, where the property of the noise is known in advance to be correlated or biased, the digital QEC can also improve the QEC performance with a single block code by considering the likelihood of the joint event of multiple qubits. By contrast, a likelihood for the GKP qubit is obtained by analog information from the measurement without any knowledge about a priori noise distribution, since the analog information  intrinsically obeys a Gaussian distribution by virtue of the GKP encoding.

The performance of the QEC under the code capacity model can be evaluated with the hashing bound of the standard deviation for the quantum capacity of the Gaussian quantum channel. In Refs.~\cite{GKP,Harri}, the hashing bound has been conjectured to be $\sim$ 0.607. We have shown using Monte Carlo method that the analog QEC with the concatenated code using the Knill's $C_{4}/C_{6}$ code~\cite{Knill} can achieve the hashing bound of the standard deviation  $\sim$ 0.607 for the quantum capacity of the Gaussian quantum channel in Ref.~\cite{KF1}. This implies that the analog QEC with the $C_{4}/C_{6}$ code provides an optimal performance against the Gaussian quantum channel. In addition, a specific method to achieve the hashing bound against the Gaussian quantum channel has not been reported except for analog QEC.

\begin{figure}[tb]
\begin{center}
 \includegraphics[angle=0, width=1\columnwidth]{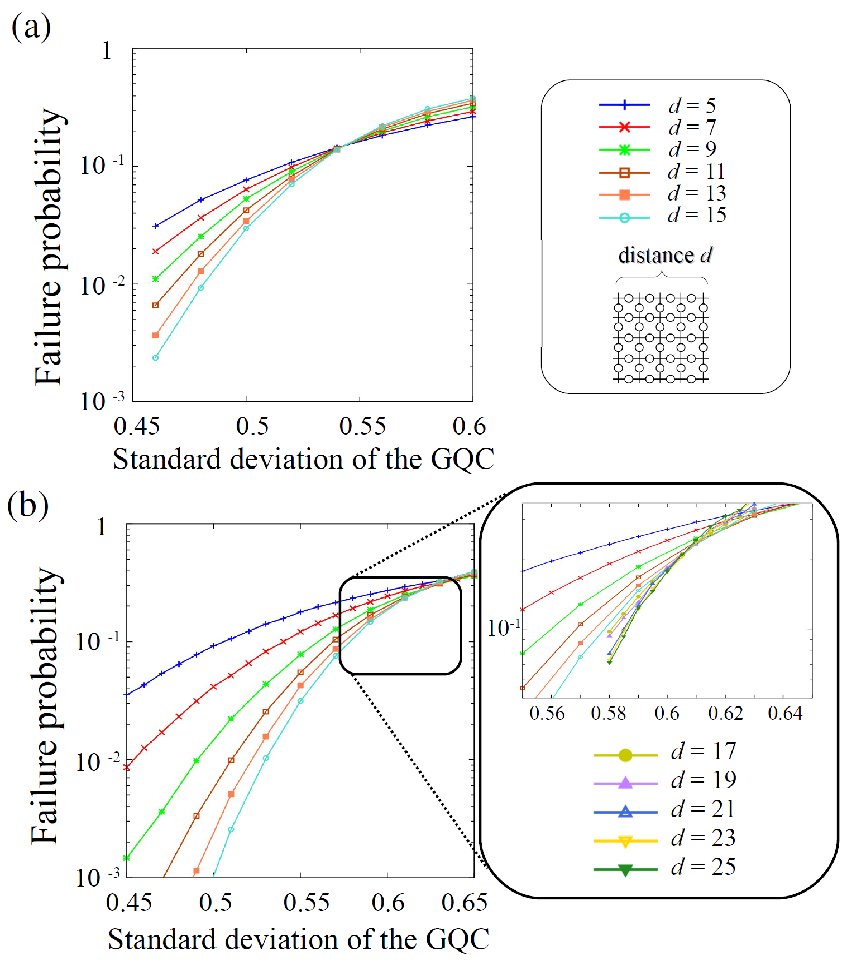}
 \end{center}
\setlength\abovecaptionskip{-10pt}
\caption{Simulation results for the logical error probabilities of the surface code with ideal syndrome measurements using (a) the digital QEC and (b) the analog QEC for several distance d which is size of the 3D cluster state. The simulation results for the digital QEC are obtained from 50000 samples. The simulation results for the analog QEC are obtained from 50000 samples (for $d=5-15)$ and 10000 samples (for $d=17-25)$ .}
 \label{fig:2}
\end{figure}
\begin{figure}[tb]
\begin{center}
 \includegraphics[angle=270, width=1.03\columnwidth]{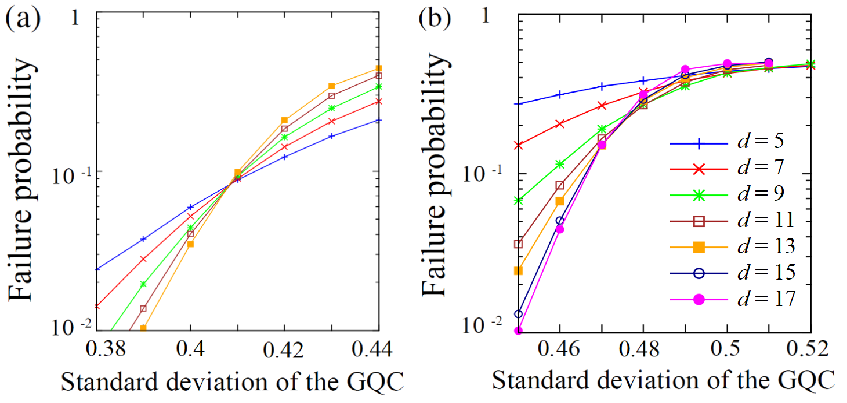}
 \end{center}
       \setlength\abovecaptionskip{-5pt}
 \caption{Simulation results for the logical error probabilities of the surface code with noisy syndrome measurements using (a) the digital QEC and (b) the analog QEC. The
simulation results for the digital QEC are obtained from 50000 samples. The simulation results for the analog QEC are obtained from 10000 samples.}
 \label{fig:3}
\end{figure}
\subsection{Analog QEC with a surface code}
While the analog QEC has been investigated by the concatenated code,  its validity on the surface code, which is one of the important candidate for scalable FTQC, is still unknown. Here we first investigate the QEC process of a surface code under the code capacity noise model, where the QEC is operated with ideal syndrome measurements. In the code capacity noise model, the surface code state consisting of the GKP qubits of an infinite squeezing suffer from the Gaussian quantum channel, which decrease the squeezing level of the GKP qubits except for the syndrome qubits. Later we will extend the analog QEC to the phenomenological noise model, where the single GKP qubits of an infinite squeezing are encoded by the surface code, and the Gaussian quantum channel decreases the squeezing level of all qubits. 

We here investigate the QEC process of a surface code with the code capacity model to verify whether analog QEC with the surface code can provide an optimal performance against the Gaussian quantum channel, since we employ topologically protected MBQC to implement FTQC. For the QEC, we employ the minimum distance decoding, which can be done efficiently by using a minimum-weight perfect matching algorithm~\cite{Ed,Kol}. In the decoding, we employ a minimum-weight perfect matching algorithm to find the most likely location of the errors according to the error syndrome. In the digital QEC, the weights are determined from the error probability a priori. Specifically, for an independent and identical error distribution the weights are chosen to be homogeneous. On the other hand, in the analog QEC, the weights are calculated by using a likelihood as 
\begin{eqnarray}
l_{\rm in}&=&-{\rm log} \left[ {f(|\Delta_{\rm m}|)}/{f(\sqrt{\pi}-|\Delta_{\rm m}|)} \right],
\end{eqnarray}
where and $l_{\rm in}$ is a likelihood for the incorrect decision. In Fig. \ref{fig:2}, the logical error probabilities are plotted as a function of the standard deviation of the GKP qubits for the code distances $d=5, 7 ,9, \cdots$. \\

To obtain the threshold value of the surface codes with the digital QEC and the analog QEC, the finite-size scaling ansatz similar to Refs.~\cite{Fujii3, Wang} was used. Specifically, the logical error probability $P_{L}=A+B(\sigma-\sigma_{th})L^{1/v}$ was used for the fitting function, where $A$, $B$, the threshold value of the Gaussian quantum channel noise $\sigma_{th}$, and $v$ are the fitting parameters. The results in Fig. \ref{fig:2} confirm that our method can reduce the logical error probability. This indicates that the analog QEC also achieves $\sim$0.607 close to the hashing bound of the quantum capacity of the Gaussian quantum channel. On the other hand, the digital QEC with only binary information achieves $\sim$0.542. This implies that the analog QEC with the surface code provides an optimal performance against the Gaussian quantum channel.  In Fig.\ref{fig:2} (b), the curves do not meet at a single point. We consider the reason is because of the finite size effect in the case of the optimal decoding for the low-distance QEC. We confirmed the finite size effect in Ref.~\cite{KF1} in the case of analog QEC. In addition, the finite size effect is distinctively seen in Fig. 1 of Ref.~\cite{Pou} in the case of the optimal decoding for the low-distance QEC.

Next, we simulate the QEC process of topologically protected MBQC with the surface code ~\cite{Rau,Rau2,Fujii4, Fujii5} under a phenomenological noise model. We here investigate the QEC process on the 3D cluster state. There are the primal and dual cubes, faces, and edges in a unit cell of the 3D cluster state. In topological QEC on the 3D clsuter state, if there is no error, the parity of each six $X$-basis measurement outcomes on the primal cube is always even. The errors are described by using a dual 1-chain and we estimate the location of errors from a set of odd parity cubes. In Fig. \ref{fig:3}, the logical error probabilities are plotted as a function of the standard deviation. The results confirm that our method can also suppress errors with the phenomenological noise model, and the threshold for the standard deviation can be improved from 0.41 to 0.47,  which corresponds to improvement of the squeezing level from 4.7 dB to 3.5 dB. Hence, analog QEC with the phenomenological noise model can reduce the required squeezing level by 1.2 dB in comparison to the digital QEC. 
\section{The 3D cluster state construction with the postselected measurement}\label{sec3}
\subsection{The accumulation of errors during the construction of the 3D cluster state}
We have investigated the QEC process of topologically protected MBQC on the 3D cluster state prepared by the infinitely squeezed GKP qubits in the previous section. In this section, we consider more realistic condition, where the 3D cluster state is constructed from the single GKP qubits of a finite squeezing level by using only the CZ gate. In the following, we refer this noise model to the correlated error model. In the construction of the 3D cluster state by using only the CZ gate, the accumulation of errors on the qubit, which causes degradation of the squeezing level, generally increases as the number of the the CZ gate on the qubit increases. The CZ gate for the GKP qubits, which corresponds to the operator exp(-$i\hat{q}_{\rm C}\hat{q}_{\rm T}$), transforms
\begin{eqnarray}
 \hat{q}_{\rm C} \to   \hat{q}_{\rm C},\hspace{26pt} \\
 \hat{p}_{\rm C} \to \hat{p}_{\rm C} - \hat{q}_{\rm T}\ ,  \\
 \hat{q}_{\rm T} \to   \hat{q}_{\rm T}, \hspace{27pt}\\
 \hat{p}_{\rm T} \to  \hat{p}_{\rm T} - \hat{q}_{\rm C}\hspace{4pt}, 
\end{eqnarray}
where $\hat{q_{\rm C}}$ ($\hat{q_{\rm T}}$) and $\hat{p_{\rm C}}$ ($\hat{p_{\rm T}}$) are the $q$ and $p$ quadrature operators of the control (target) qubit, respectively. 

We here consider the error propagation caused by the CZ gate operation. The CZ gate operation displaces the deviation for the $q$ and $p$ quadrature as
\begin{eqnarray}
\overline{\Delta}_{\rm {\it q},C} \to  \overline{\Delta}_{\rm {\it q},C} , \hspace{30pt}\\
 \overline{\Delta}_{\rm {\it p},C} \to \overline{\Delta}_{\rm {\it p},C}- \overline{\Delta}_{\rm {\it q},T},  \\
\overline{\Delta}_{\rm {\it q},T} \to  \overline{\Delta}_{\rm {\it q},T} ,  \hspace{30pt}\\
\overline{\Delta}_{\rm {\it p},T} \to \overline{\Delta}_{\rm {\it p},T}- \overline{\Delta}_{\rm {\it q},C}, 
\end{eqnarray}
where $\overline{\Delta}_{\rm {\it q},C}$ ($\overline{\Delta}_{\rm {\it q},T}$) and $\overline{\Delta}_{\rm {\it p},C}$ ($\overline{\Delta}_{\rm {\it p},T}$)  are true deviation values in the $q$ and $p$ quadrature of the control and target qubit, respectively. Since the true deviation obeys Gaussian distribution and takes a value randomly and independently, the variance of the control qubit and target qubit in $p$ quadrature changes as
\begin{eqnarray}
{\sigma}^2_{p,{\rm C}}   \to  {\sigma}^2_{p,{\rm C}}+ {\sigma}^2_{q,{\rm T}} ,  \\
{\sigma}^2_{p, {\rm T}} \to  {\sigma}^2_{p,{\rm T}}+{\sigma}^2_{q,{\rm C}} ,
\end{eqnarray}
where ${\sigma}^2_{q,{\rm C}} ({\sigma}^2_{q,{\rm T}})$ and  ${\sigma}^2_{p,{\rm C}} ({\sigma}^2_{p,{\rm T}})$ are the variance of the control and target qubit in the $q$ and $p$ quadrature, respectively. On the other hand, the variance in the $q$ quadrature does not change. Therefore, the CZ gate increases the probability of misidentifying the bit value in $p$ quadrature. In this work, we define the error propagation caused by the CZ gate operation as the correlated error. Assuming that the variance of the single GKP qubit is $\sigma^2$, if the 3D cluster state is prepared straightforwardly by only the CZ gates, the variance of the qubits in the $p$ quadrature become $5\sigma^2$, since the qubit of the 3D cluster state is generated by using the CZ gates between four neighboring qubits. This deteriorates the required squeezing level for the surface code from 4.7 dB under the phenomenological noise model to 13.7 dB under the correlated noise model. In the following, we propose the postselected measurement to avoid the accumulation of the errors, which allows us to achieve a high threshold.
\begin{figure}[t]
 \includegraphics[angle=270, scale=1.05]{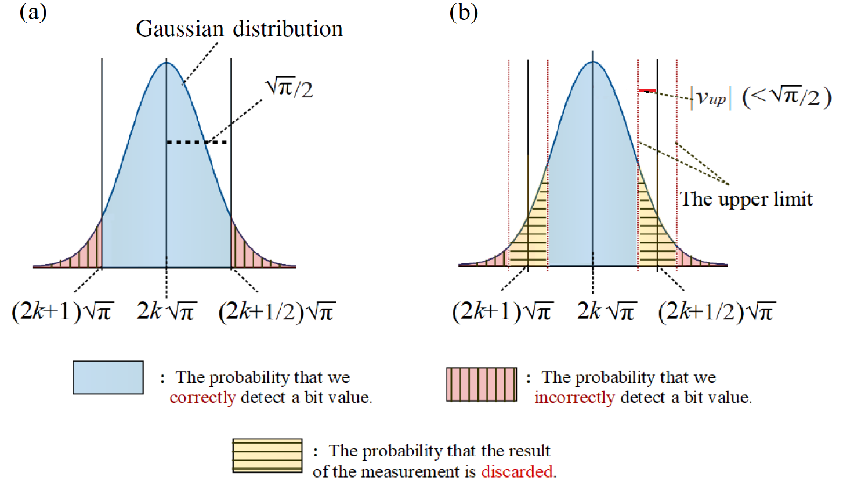} 
     \setlength\abovecaptionskip{0pt}
\caption{Introduction of the postselected measurement. (a) The conventional measurement of the GKP qubit, where the Gaussian distribution followed by the deviation of the GKP qubit that has variance $\sigma^{2}$. The  plain (blue) region and the region with vertical (red) line represent the different code word $(k-1)$ mod 2 and $(k+1)$ mod 2, respectively. The vertical line regions correspond to the probability of incorrect decision of the bit value. (b) The postselected measurement. The shown dot line represents a upper limit $v_{\rm up}$. The horizontal line areas show the probability that the results of the measurement is discarded by introducing $v_{\rm up}$. The vertical line areas show the probability that our method fails.}
\label{fig:4}
\end{figure}
\begin{figure}[t]
    \includegraphics[angle=0, width=0.7\columnwidth]{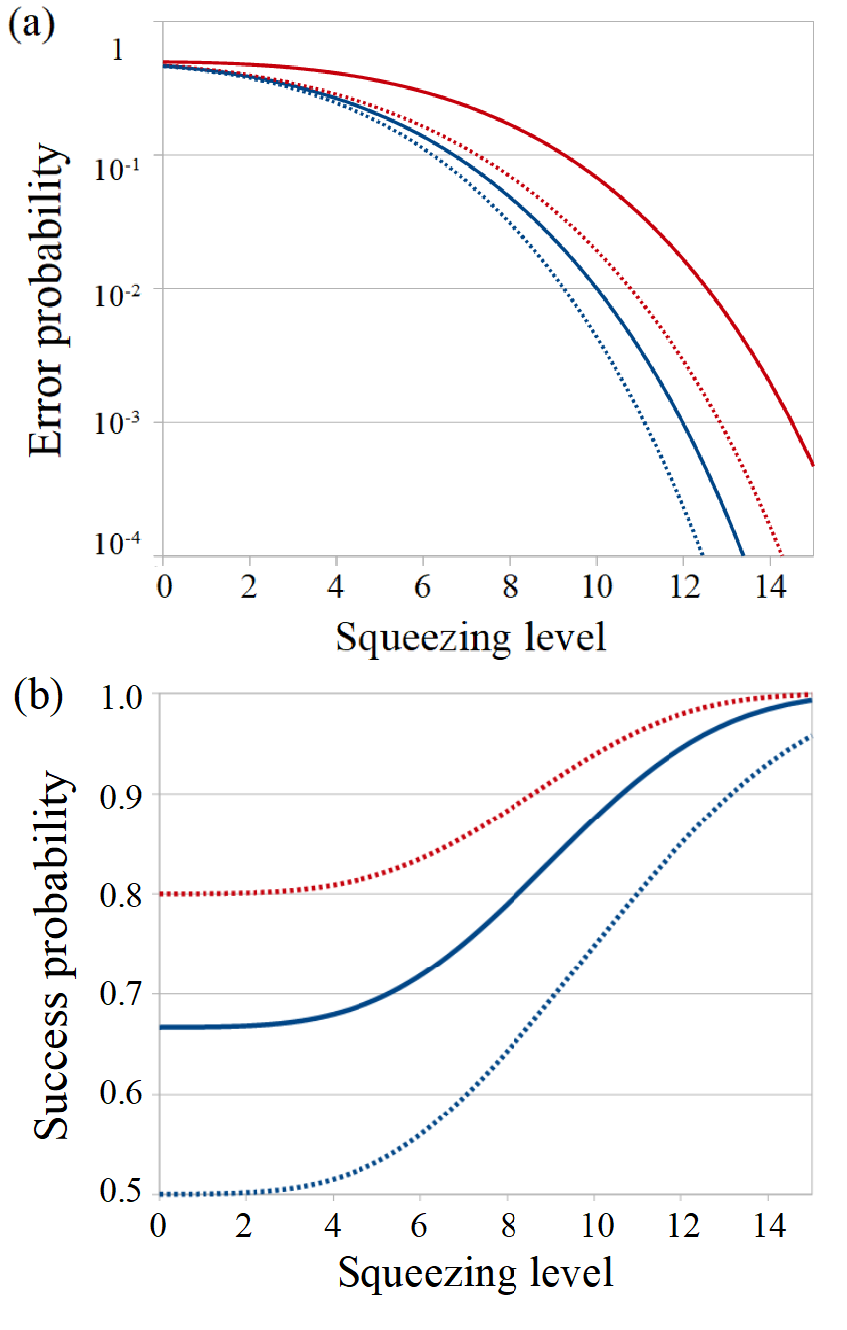}  
          \setlength\abovecaptionskip{-3pt}
    \caption{The error probabilities of the postselected measurement $E_{\rm post}$ and the success probabilities of the postselected measurement $P_{\rm Suc}$ on the qubit of the variance $3\sigma^2$. (a) The error probabilities with the method using the only CZ gate and our method using the postselected measurement for the upper limit $v_{up}=0$ (red solid), $v_{up}=\sqrt{\pi}/10$ (red dashed), $v_{up}=\sqrt{\pi}/6$ (blue solid), and $v_{up}=\sqrt{\pi}/4$ (blue dashed), respectively. (b) The success probability for our method.  The squeezing level is equal to $-10 {\rm log}_{10}6\sigma^{2}$. }
    \label{fig:5}
\end{figure}
\subsection{The postselected measurement}
We propose the postselected measurement that utilizes analog information, and explain a method to generate an entanglement between the qubits by using the postselected measurement, avoiding the accumulation of errors during the construction process. Following Ref.~\cite{Brow}, we call this entanglement generation with the postselected measurement as the fusion gate.
As we mentioned, in the measurement of the GKP qubit, we make a decision on the bit value $k (= 0, 1)$ from the measurement outcome of the GKP qubit $q_{\rm m}= q_{k}+ \Delta_{\rm m}$. The conventional decision sets an upper limit for $|\Delta_{m}|$ at $\sqrt{\pi}/2$, and assigns the bit value $k$ = $(2 t + k)\sqrt{\pi} $. The decision is correct as long as the amplitude of the true deviation $|\overline{\Delta}|$ falls between 0 and $\sqrt{\pi}/2$. The probability to obtain the correct bit value is thus given by $p_{\rm corr}$ in Eq. (\ref{eq:7}). The proposed decision sets an upper limit at $v_{\rm up}(<\sqrt{\pi}/2)$  to give the maximum deviation that will not cause incorrect measurement of the bit value as shown in Fig. \ref{fig:4}. If the above condition $|\Delta_{\rm m}| < v_{\rm up}$ is not satisfied, we discard the result. Since the measurement error occurs when $|\bar{\Delta}|$ exceeds $|\sqrt{\pi}/2+v_{\rm up}|$, the error probability decreases as increasing $v_{\rm up}$ at the cost of the success probability of the measurement. The probability to obtain the correct bit value with the postselected measurement $P_{\rm post}$ is equal to ${P_{\rm post}^{cor}}/({P_{\rm post}^{cor}}+{P_{\rm post}^{in}})$, where ${P_{\rm post}^{cor}}$ is the probability that the true deviation $|\bar{\Delta}|$ falls in the correct area, and ${P_{\rm post}^{in}}$ is the probability that the true deviation $|\bar{\Delta}|$ falls in the incorrect area. ${P_{\rm post}^{cor}}$ and ${P_{\rm post}^{in}}$ for the GKP qubit of the variance $\sigma^2$ are given by
\begin{equation}
P_{\rm post}^{cor}= \sum_{k=-\infty}^{+\infty} \int_{2k\sqrt{\pi}-\frac{\sqrt{\pi}}{2}+v_{up}}^{2k\sqrt{\pi}+\frac{\sqrt{\pi}}{2}-v_{up}} dx \frac{1}{\sqrt{2\pi {\sigma}^2}}\mathrm{e}^{-\frac{x^2}{{2{\sigma}^2}}}
\end{equation}
and 
\begin{equation}
P_{\rm post}^{in}= \sum_{k=-\infty}^{+\infty} \int_{(2k+1)\sqrt{\pi}-\frac{\sqrt{\pi}}{2}+v_{up}}^{(2k+1)\sqrt{\pi}+\frac{\sqrt{\pi}}{2}-v_{up}} dx \frac{1}{\sqrt{2\pi {\sigma}^2}}\mathrm{e}^{-\frac{x^2}{{2{\sigma}^2}}}.
\end{equation}
\begin{figure*}[t]
    \includegraphics[angle=270, width=2.0\columnwidth]{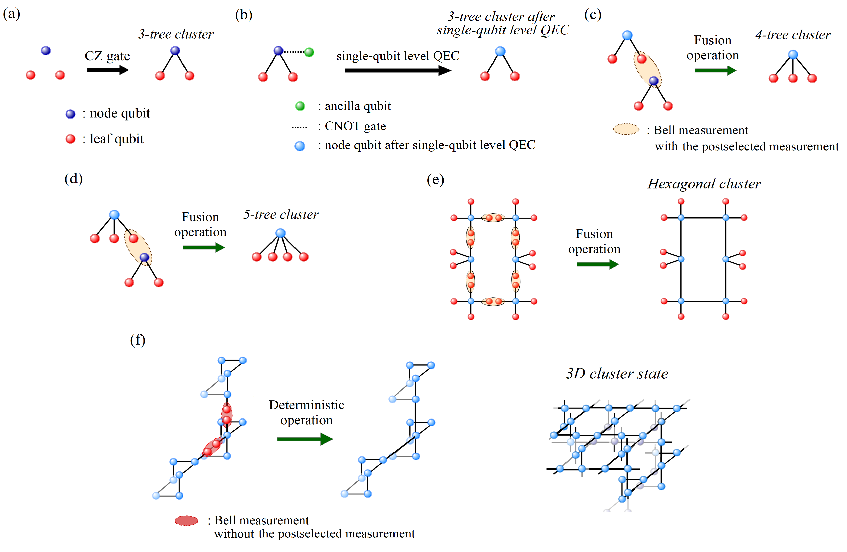} 
    \setlength\abovecaptionskip{15pt}
    \caption{The 3D cluster state construction. (a) The preparation of the 3-tree cluster state by using the CZ gate. (b) The single-qubit level QEC using the additional ancilla qubit with the postselected measurement. (c)--(e) The construction of the hexagonal cluster state from the 3-tree qubit with the postselected measurement. (f) The construction of the 3D cluster state from the hexagonal cluster states, where the entanglement is generated between the neighboring hexagonal cluster states without the postselected measurement.
}
\label{fig:6}
\end{figure*}
In Fig. \ref{fig:5}, we plot the probability to misidentify the bit value with the postselected measurement $E_{\rm post}=1-P_{\rm post}$ and the success probability of the post-selection $P_{\rm Suc}={P_{\rm post}^{cor}}+ {P_{\rm post}^{in}}$ as a function of the squeezing level for several $v_{\rm up}$. As an example, we described the  measurement on the qubit of the variance $3\sigma^2$, which is frequently occurred in the Bell measurement during the construction process. Fig. \ref{fig:5} shows that both the error probability $E_{\rm post}$ and the success probability $P_{\rm Suc}$ decrease. 
In our method, we apply the postselected measurement with $v_{up}=2\sqrt{\pi}/5 $ to the 3D cluster states construction to prevent the deviation of the GKP qubit from propagating the qubit-level error derived from the fusion gate. Because of the postselected measurement, the operation such as the fusion gate becomes nondeterministic. This will be also handled by the so called divide and conquer approach~\cite{Niel, Daw} below.
\subsection{The 3D cluster state construction}
We explain how to apply the postselected measurement to prevent the squeezing level from decreasing during the construction of the 3D cluster state. Hereafter, we omit  \textquotedblleft GKP\textquotedblright\ of the GKP qubit and simply call it as a qubit. In our method, there are four steps. In step 1, we prepare a node qubit and two leaf qubits of the variance ${\sigma}^2$ in the $q$ and $p$ quadrature (Fig. \ref{fig:6} (a)). By using the CZ gate we obtain 3-tree cluster state composed of a node qubit and two leaf qubits, where the the variance of the node and leaf qubits in the $p$ quadrature increase from ${\sigma}^2$ to $3{\sigma}^2$ and $2{\sigma}^2$ , respectively. On the other hand, the variance of the node and leaf qubits in the $q$ quadrature keep the variance ${\sigma}^2$.

In step 2, we operate the single-qubit level QEC ~\cite{GKP,Gla,Steane} by using the CNOT gate with the postselected measurement (Fig. \ref{fig:6} (b)). In this single-qubit level QEC, the additional single ancilla qubit is entangled with the node qubit by using the CNOT gate, assuming the node qubit is the target qubit. The ancilla qubit is prepared in the state $\ket {\widetilde{0}}$ to prevent us from identifying the bit value of the node qubit. The CNOT gate, which corresponds to the operator exp(-$i\hat{q}_{\rm C}\hat{p}_{\rm T}$), transforms
\begin{eqnarray}
 \hat{q}_{\rm C} \to   \hat{q}_{\rm C},\hspace{26pt} \\
 \hat{p}_{\rm C} \to \hat{p}_{\rm C} - \hat{p}_{\rm T}\ ,  \\
 \hat{q}_{\rm T} \to   \hat{q}_{\rm T}+ \hat{q}_{\rm C},\hspace{4pt} \\
 \hat{p}_{\rm T} \to  \hat{p}_{\rm T},  \hspace{27pt}
\end{eqnarray}
Regarding the deviation, the CNOT gate operation displaces the deviation for the $q$ and $p$ quadrature as
\begin{eqnarray}
\overline{\Delta}_{\rm {\it q},C} \to  \overline{\Delta}_{\rm {\it q},C} ,  \hspace{30pt}\\
 \overline{\Delta}_{\rm {\it p},C} \to \overline{\Delta}_{\rm {\it p},C}- \overline{\Delta}_{\rm {\it p},T}, \\
\overline{\Delta}_{\rm {\it q},T} \to  \overline{\Delta}_{\rm {\it q},T}+ \overline{\Delta}_{\rm {\it q},C} ,  \hspace{1pt}\\
\overline{\Delta}_{\rm {\it p},T} \to \overline{\Delta}_{\rm {\it p},T}.\hspace{31pt}
\end{eqnarray}
After the CNOT gate, we measure the ancilla qubit in the $p$ quadrature and obtain the deviation of the ancilla qubit ${\Delta}_{\rm m{\it p}, a}$. In the single-qubit level QEC, if $|{\Delta}_{\rm  m{\it p}, a}| = |\overline{\Delta}_{\rm {\it p},a}- \overline{\Delta}_{\rm {\it p},n}|$ is less than $\sqrt{\pi}/2$, the true deviation value of the node qubit in the $p$ quadrature changes from $\overline{\Delta}_{\rm {\it p},n}$ to $ \overline{\Delta}_{\rm {\it p},a}$ after the displacement operation, which displaces $\overline{\Delta}_{\rm {\it p},n}$ by ${\Delta}_{\rm  m{\it p}, a}(=\overline{\Delta}_{\rm {\it p},a}- \overline{\Delta}_{\rm {\it p},n})$. On the other hand, if $ |\overline{\Delta}_{\rm {\it p},a}- \overline{\Delta}_{\rm {\it p},n}|$ is more than $\sqrt{\pi}/2$, the bit error in the $p$ quadrature occurs after the displacement operation. This error can be reduced by the postselected measurement on the ancilla defined as follows: if $|{\Delta}_{\rm  m{\it p}, a}|$ is less than $v_{\rm up}$, we then operate the displacement to the node qubit in the $p$ quadrature by ${\Delta}_{\rm m{\it p}, a}$. Otherwise, we discard the resultant 3-tree cluster state and restart the procedure from step 1. The error probability of the single-qubit level QEC $E_{\rm SQE}$ is given by $E_{\rm post}$ defined in the previous section with the variance of $4\sigma^2$, since after the CNOT gate the true deviation of the ancilla qubit in the $p$ quadrature $|\overline{\Delta}_{\rm {\it p},a}- \overline{\Delta}_{\rm {\it p},n}|$ obeys Gaussian distribution with the variance $4{\sigma}^2$, where $4\sigma^2$ comes from the node qubit and $\sigma^2$ from the ancilla qubit. To summarize, the single-qubit level QEC can reduce the variance of the node qubit in the $p$ quadrature from $3{\sigma}^2$ to ${\sigma}^2$, since $\overline{\Delta}_{\rm {\it p},a}$ and $\overline{\Delta}_{\rm {\it p},n}$ obey Gaussian distributions with the variances $3\sigma^2$ and $\sigma^2$, respectively. The variance of the node qubit in the $q$ quadrature after the single-qubit level QEC increases from ${\sigma}^2$ to $2{\sigma}^2$, since the true deviation $\overline{\Delta}_{\rm {\it p},n}+ \overline{\Delta}_{\rm {\it q},a}$ obeys Gaussian distribution with the variance $2{\sigma}^2$, where the $\overline{\Delta}_{\rm {\it p},n}$ and $\overline{\Delta}_{\rm {\it q},a}$ are the true deviation of the node qubit and the ancilla qubit, respectively. This increase in the variance in $q$ quadrature has no effect on the threshold value, whereas the unheralded error in the $p$ quadrature affects it. 

In step 3, we increase the number of the leaf qubits of the tree cluster state by using the fusion gate with the postselected measurement. The fusion gate can avoid the deviation of the qubit from increasing and the postselected measurement can prevent the qubit-level error from propagating during constructing the 6-tree cluster state, which we call the hexagonal cluster state. We describe the construction of the 4-tree cluster state in detail as follows. By using the fusion gate, we construct the 4-tree cluster state from the two 3-tree cluster states, one of which is corrected by the single-qubit level QEC and the other is uncorrected (Fig. \ref{fig:6} (c)). In the fusion gate, the Bell measurement with the postselected measurement is implemented by beam splitter coupling and homodyne measurement. Then feedforward is operated according to the homodyne measurement outcomes on the leaf and the node qubits, respectively. If the misidentification of the bit value of the leaf or node qubits occurs, the feedforward operation propagates the qubit-level error in the 4-tree cluster. The probabilities to misidentify the bit value of the leaf and node qubits are the probabilities to misidentify the bit value of the qubit of the variances $3\sigma^2$ and $4\sigma^2$, respectively. This unheralded qubit-level error can be reduced by using the postselected measurement. We define the unheralded errors on the leaf qubits and node qubits with the postselected measurement as $E_{\rm post}( 3{\sigma}^2)$ and $E_{\rm post}( 4{\sigma}^2)$, respectively. The error probabilities $E_{\rm post}( 3{\sigma}^2)$ and $E_{\rm post}( 4{\sigma}^2)$ are given by $E_{\rm post}$ defined in the previous section with the variance of $3\sigma^2$ and $4\sigma^2$, respectively.

To evaluate the variances of the leaf and node, we describe the process of the beam splitter coupling in the following. The 50:50 beam splitter coupling between the leaf qubit of the 3-tree cluster state after the single-qubit level QEC and the node qubit of the 3-tree cluster state without the single-qubit level QEC transforms the variables of the leaf and node qubits in the $q$ and $p$ quadrature as
\begin{eqnarray}
\hat{q}_{\rm leaf}  &\to &     (\hat{q}_{\rm leaf}+\hat{p}_{\rm node})/\sqrt{2},  \\ 
\hat{p}_{\rm leaf} &\to &  (\hat{p}_{\rm leaf}+\hat{q}_{\rm node})/\sqrt{2},\\
\hat{q}_{\rm node}  &\to &  (\hat{q}_{\rm leaf}-\hat{p}_{\rm node})/\sqrt{2},  \\ 
\hat{p}_{\rm node}  &\to &  (\hat{p}_{\rm leaf}-\hat{q}_{\rm node})/\sqrt{2}, 
\end{eqnarray}
where $ \hat{q}_{\rm leaf} $ ($ \hat{q}_{\rm node} $) and  $\hat{p}_{\rm leaf} $ ($\hat{p}_{\rm node}$) the variables of the leaf (node) qubit in the $q$ and $p$ quadrature, respectively. After the coupling, the variances of the leaf qubit in the $q$ and $p$ quadrature changes as $\sigma^2 \to 2\sigma^2$ and $2{\sigma^2} \to 3{\sigma^2}/2$, respectively. The variances of the node qubit in the $q$ and $p$ quadrature changes as $\sigma^2 \to 3{\sigma^2}/2$ and $3{\sigma^2}\to 2\sigma^2$, respectively. After the homodyne measurement on the leaf and node qubit in the $p$ quadrature, the measurement outcome of the leaf and node qubit in the $p$ quadrature are rescaled by multiplying the measurement outcome by $\sqrt{2}$ in a post-process as $(p_{\rm leaf}+{q}_{\rm node})/\sqrt{2} \to  p_{\rm leaf}+{q}_{\rm node}$ and $(p_{\rm leaf}-{q}_{\rm node})/\sqrt{2} \to p_{\rm leaf}-{q}_{\rm node}$, respectively. The variances of the leaf and node qubits in the $p$ quadrature changes as $3{\sigma^2}/2 \to 3{\sigma^2}$ and $2{\sigma^2}\to 4\sigma^2$, respectively. Therefore, the probabilities to misidentify the bit value of the leaf and node qubits in the $p$ quadrature are the probabilities to misidentify the bit value of the qubit of the variances $3\sigma^2$ and $4\sigma^2$, respectively. 

We can reduce the misidentifying error probabilities occurred in the construction of the hexagonal cluster state in the same way. We generate the 5-tree cluster state from the 3-tree cluster states and the 4-tree cluster state by using fusion gate with the postselected measurement on the leaf qubit of the 3-tree cluster state and the node qubit of the 4-tree cluster state with the postselected measurement (Fig. \ref{fig:6} (d)). Finally, we construct the hexagonal cluster state from the six 5-tree cluster states with the postselected measurement on the Bell measurement between leaf qubits (Fig. \ref{fig:6} (e)). 

In step 4,  we generate the 3D cluster state deterministically. Hence, the postselected  measurement can not be used and the 3D cluster state is generated from the hexagonal cluster states by using the fusion gate with the postselected measurement between the leaf qubits of the neighboring hexagonal cluster states without the postselected measurement (Fig. \ref{fig:6} (f)). In this step, the unheralded error, which corresponds to the probability to misidentify the bit value of the qubit of the variance $3{\sigma}^2$, accumulates on the node qubits. We define this error probability as $E_{{\rm Bell}}$. We can eventually obtain the 3D cluster state composed of the node qubits whose variance and squeezing level in the $p$ quadrature are ${\sigma}^2$ and $-10 {\rm log}_{10}2\sigma^{2}$, respectively. 

By contrast, the conventional method, where the fusion gate with the postselected measurement is not used and the 3D cluster state is generated by using only the CZ gate between neighboring nodes, yields the variance $5{\sigma}^2$ and the squeezing level $-10 {\rm log}_{10}10\sigma^{2}$ of node qubits in the $p$ quadrature, respectively. Therefore, the single-qubit level QEC and the fusion gate with postselected measurement can avoid the degradation of the squeezing level during the construction of the 3D cluster state. This relaxes the requirement on the squeezing level of the initial single qubit considerably as will be calculated in the next section.
\section{Threshold calculation for topologically protected MBQC}\label{sec4}
In this section, we calculate the threshold value for the 3D cluster state prepared by using the postselected measurement. In this calculation, it is assumed that the 3D cluster state is prepared using the proposed method with the qubit of the variances finite value $\sigma^2$ in $q$ and $p$ quadrature, that is, the initial variances of the qubit before the CZ gate in Fig. \ref{fig:6} (a) are $\sigma^2$ in $q$ and $p$ quadrature. 

We define the unheralded error probability $E_{\rm tot}$ per one node qubit of the 3D cluster state in the $p$ quadrature. The $E_{\rm tot}$ results from four causes, which correspond to the error originated from the node qubit itself, the unheralded errors during postselected measurements in steps 2 and 3, and the error during the deterministic fusion gate in step 4. The unheralded error of the node qubit itself $E_{\rm node}$ occurs, when the magnitude of the true deviation value of the node qubit is more than $\sqrt{\pi}/2$. The error probability $E_{\rm node}$ is given by $E_{\rm post}$ with the variance in $p$ quadrature $\sigma^2$. The unheralded error probability in the single-qubit level QEC (step 2)  $E_{\rm SQE}$ is given by $E_{\rm post}$ with the variance in $p$ quadrature $4\sigma^2$. The unheralded errors during postselected measurements occurs in the two processes of step 3. One is in the 4- and 5-tree cluster states construction by using the Bell measurement shown in Fig. \ref{fig:6} (c) and (d). The probabilities of misidentifying the bit value on the node qubit in the Bell measurement are both $E_{\rm post}(4\sigma^2)$ given by $E_{\rm post}$ with the variance in $p$ quadrature $4\sigma^2$. The other unheralded error process of the postselected measurement in step 3 is the bit value misidentification on the leaf qubits by using the fusion gate in Fig. \ref{fig:6} (c)-(e). The error probability $E_{\rm post}(3\sigma^2)$ is given by $E_{\rm post}$ with the variance in $p$ quadrature $3\sigma^2$. 
The measurement error in the deterministic entanglement generation between neighboring node qubits occurs in the Bell measurement between the leaf qubits of the hexagonal cluster states without the postselected measurement (Fig. \ref{fig:6} (f)). This unheralded error probability $E_{\rm Bell}$ corresponds to the probability of misidentifying the bit value on the qubit of the variance 3${\sigma}^2$ without the postselected measurement. This process requires two Bell measurements per one node qubit as shown in Fig. \ref{fig:6} (f)

For simplicity, we firstly calculate the $E_{\rm tot}$ in the leading order. Later we will take more detailed calculation by the simulation of the QEC for topologically protected MBQC by using the minimum-weight perfect matching algorithm. 

The error probability $E_{\rm tot}$ in the leading order can be obtained as 
\begin{eqnarray}
 E_{{\rm tot}} &=& E_{{\rm node}}({\sigma}^2 )+  E_{\rm SQE}+6{\times}E_{\rm post}( 3{\sigma}^2)  \nonumber  \\
&+& 2 \times E_{\rm post}({ 4\sigma}^2) + 2\times E_{{\rm Bell}}. 
\end{eqnarray}
We estimated the required squeezing level for CV-FTQC in the leading order as follows. Let us first consider the case without analog QEC. By virtue of the postselected measurements, the correlated error probability on the 3D cluster state is now very small and can be neglected safely. In fact, for around 10 dB squeezing with $v_{up}=2\sqrt{\pi}/5$, the unheralded error probability $E_{\rm post}( 3{\sigma}^2)$ and $E_{\rm post}( 4{\sigma}^2)$ is order of $10^{-5}$ and  $10^{-4}$, which is much smaller than the unheralded error probability $E_{{\rm Bell}}$ of about 1.5 $\%$~\cite{Note3}. Hence, since we can ignore the correlated errors on the 3D cluster state, the error probability $E_{\rm tot}$ can be fairly well characterized by the single parameter $\sigma^2$ under the phenomenological noise model, where the required squeezing level for topologically protected MBQC is 2.9-3.3$\%$~\cite{Wang,Ohno}. We define the required squeezing level as the squeezing level that provides $E_{tot}=3.0\%$, and the numerical calculation in the leading order without analog QEC  yields the required squeezing level of 10.5 dB with $v_{up}=2\sqrt{\pi}/5$. We can further improve the tolerable standard deviation by using analog QEC. In Sec. \ref{sec2} we numerically simulated the improvement of the topologically protected MBQC performance in the analog QEC with the phenomenological noise model, and obtained the improvement on the required squeezing level by 1.2 dB in comparison to the digital QEC as shown in Fig. \ref{fig:3}. Hence, we can obtain the required squeezing level 9.3 dB in the leading order with analog QEC. 
\begin{figure}[t]
    \includegraphics[angle=0, width=1\columnwidth]{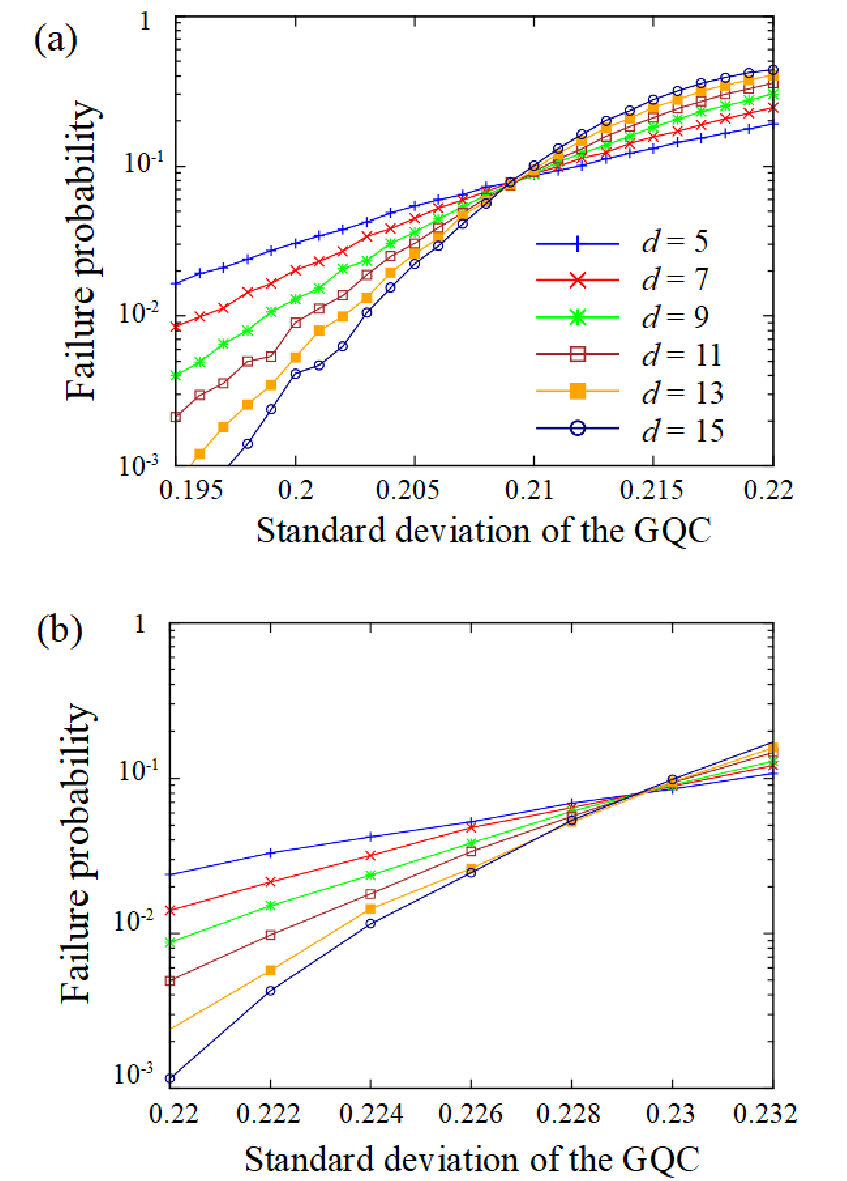} 
              \setlength\abovecaptionskip{-3pt}
    \caption{Simulation results for the logical error probabilities of the surface code by using the 3D cluster state prepared by the proposed method for $v_{up}=2\sqrt{\pi}/5$ with noisy syndrome measurements using (a) the digital QEC and (b) the analog QEC, respectively.  The simulation results for the digital QEC are obtained from 50000 samples. The simulation results for the analog QEC are obtained from 10000 samples.}
    \label{fig:7}
\end{figure}

To proceed to the detailed calculation of  $E_{\rm tot}$, we simulate the QEC for topologically protected MBQC by using the minimum-weight perfect matching algorithm. In Fig. \ref{fig:7}, the logical error probabilities are plotted as a function of the standard deviation. The results confirm that our method can also suppress errors with the independent error model, and the threshold for the standard deviation can be improved from 0.208 to 0.228, which corresponds to squeezing level from 10.6 dB to 9.8 dB. In the numerical calculation, we set the upper limit $v_{\rm up}$ to $2\sqrt{\pi}/5$ in order to adopt the independent error model. Therefore, CV-FTQC with analog QEC and the postselected measurement  can improve the required squeezing level for topologically protected MBQC by 6.2 dB in comparison to the existing scheme for CV-FTQC ~\cite{Meni,Note}.

Finally, we examined the resource required per node qubit composing the 3D cluster states, namely the average number of the 3-tree cluster states to construct the hexagonal cluster state. The average number of the 3-tree cluster states to construct the 5-tree cluster state can be counted as $R_{\rm 5tree}=(1/P_{\rm SQEC}+2)/{P^{2}_{\rm Bell}}$, where $P_{\rm SQEC}$ and $P_{\rm Bell}$ are the success probability of the single-qubit level QEC and of the Bell measurement with postselected measurement, respectively. $P_{\rm SQEC}$ and $P_{\rm Bell}$ are calculated as $P_{\rm Suc}(4\sigma^2)$ and  $P_{\rm Suc}(3\sigma^2) \times P_{\rm Suc}(4\sigma^2)$, respectively, where $P_{\rm Suc}(3\sigma^2)$ ($P_{\rm Suc}(4\sigma^2)$) is the success probability of the postselected measurement on the qubit of the variance is $3\sigma^2$ ($4\sigma^2$). Similarly, the average number of the 3-tree cluster states to construct the hexagonal cluster state can be counted as $R_{\rm Hexa}=(1/{P^{2}_{\rm Bell-II}}+1)\times (2/{P^{3}_{\rm Bell-II}})$, where  ${P}_{\rm Bell-II}$ is equal to ${P_{\rm Suc}}^2(3\sigma^2)$. Therefore, the resources per the hexagonal cluster states $R_{\rm Hexa}$ with $v_{up}=2\sqrt{\pi}/5$ can be estimated as $9.2\times 10^{6}$ to achieve the required squeezing level 9.8 dB, since $P_{\rm Suc}(3\sigma^2)$ and $P_{\rm Suc}(4\sigma^2)$ with the squeezing level 9.8 dB are 34.6 $\%$ and 30.2 $\%$, respectively. 
\section{Discussion and conclusion}\label{sec5}
In this work, we have proposed a high-threshold FTQC to alleviate the required squeezing level for CV-FTQC by harnessing analog information contained in the GKP qubits. The proposed method consists of applying analog QEC to the surface code and constructing the cluster state for the topologically protected MBQC with a low error accumulation by using the postselected measurement. We have numerically shown that the required squeezing level can be improved to less than 10 dB with analog QEC on the 3D cluster states prepared by using the fusion gate with the postselected measurement. Furthermore, we have numerically investigated validity of analog QEC for the surface code against the Gaussian quantum channel with ideal syndrome measurements. The numerical results have shown the analog QEC also achieves $\sim$0.607 close to the hashing bound of the quantum capacity of the Gaussian quantum channel. To the best of our knowledge, no method to provide the optimal performance has been reported except for analog QEC.

The use of analog information has been developed in classical error correction against the disturbance such as an additive white Gaussian noise~\cite{Kai} and identified as an important tool for qubit readout~\cite{Dan1,Dan2,Note4}. However, use of analog information has been left unexploited to improve the QEC performance, where superposition of the encoded qubits need to be maintained. In this work, we have shown that analog QEC can improve the QEC performance to implement a high-threshold FTQC.

To generate the GKP qubit, several methods have been proposed~\cite{Vas,Ter,Meni2,Ter2,Tra,Pet,Pir,Bar}. In particular, a promising proposal~\cite{Ter} recently exists to prepare a good GKP qubit in circuit quantum electrodynamics with the squeezing level around 10 dB~\cite{Cas}. This suggests that the GKP qubit with the squeezing level around 10 dB will be able to generate within the reach of near-term experimental set-up. Our method can achieve this experimental requirement for the squeezing level, taking a step closer to the realization of large-scale quantum computation. Hence, this is the novel application of the analog information for the practical large scale MBQC. 

We would like to mention the physical implementation for CV-FTQC with our scheme. In our method, although the fusion gate is nondeterministic, there are a number of studies for the architecture that deal with topologically protected MBQC with nondeterministic fusion gate~\cite{Seg1,Seg2,Li,Fujii6,Mon}. Our method can be implemented by these architecture straightforwardly. Considering these architecture, it is assumed that the hexagonal cluster state is prepared from the 3-tree cluster state by a purely linear optical network, composed of beam splitter coupling, an optical switch, and so on, while it is assumed that we can use the on-demand sources of the 3-tree cluster state. 

Furthermore, analog QEC and the postselected measurement can be not limited to the GKP qubit but widely applicable MBQC using various QEC codes~\cite{Bom,Ste,Bac}, and is a versatile tool for improvement of the QEC performance and the decision error of the bit value, which can incorporate with GKP qubit, cat code, and other various codes used to digitize CV states. Hence, we believe this work will open up a new approach to QEC with digitized CV states, which will be indispensable to construct CV-FTQC. 
\section*{Acknowledgements}
This work was funded by ImPACT Program of Council for Science, Technology and Innovation (Cabinet Office, Government of Japan). K. Fujii is supported by KAKENHI No. 16H02211, JST PRESTO JPMJPR1668, JST ERATO JPM- JER1601, and JST CREST JPMJCR1673.

\begin{thebibliography}{99}
\bibitem{Note}
In Ref.~\cite{Meni}, the CZ gate is performed by the measurement on squeezed vacuum states on a 2D cluster state with ancilla GKP qubits used to implement the single-qubit level QEC. Then, the fault-tolerant threshold of the squeezing level is obtained from the error probability of the the single-qubit level QEC after the CZ gate combining with the conventional fault-tolerant schemes. While in Ref.~\cite{Meni} the threshold value $10^{-6}$~\cite{Knill2,Pre,Kitaev} to calculate the required squeezing level 20.5 dB is employed, here we use the threshold value $0.67\%$ obtained in Ref.~\cite{Rau3} to calculate the required squeezing level 16.0 dB to directly compare the existing scheme with the proposed scheme, where the analog QEC is applied. We should note that it is unfair to directly compare the present result with the required squeezing level 20.5 dB in Ref.~\cite{Meni}, since optimization of the squeezing level would not be not the main scope of Ref.~\cite{Meni}.
\end{thebibliography}

\end{document}